\documentclass[10pt,aps,prb,twocolumn,superscriptaddress]{revtex4-2}
\usepackage[utf8]{inputenc}
\usepackage{graphics}
\usepackage{epsfig}
\usepackage{dcolumn}
\usepackage{bm}
\usepackage{amsmath}
\usepackage{dsfont} 
 \usepackage{mathtools}
 \usepackage{amsfonts}
\usepackage{latexsym}
\usepackage{amssymb}
\usepackage{color}
\usepackage{hyperref}

\hypersetup{
    colorlinks=true,
    linkcolor=blue,
    filecolor=magenta,      
    urlcolor=cyan,
}
\usepackage[normalem]{ulem}
 \usepackage{bbm}
\usepackage{bbold}

\usepackage{amsfonts}

 \newcommand{\g}{\bf}
\usepackage{braket}

\begin{document} 
\title{Topological  Mott insulator in  the odd-integer filled Anderson lattice model\\ with Hatsugai-Kohmoto interactions}

\author{Krystian Jab\l{}onowski}   
  \affiliation{International Research Centre MagTop, Institute of
    Physics, Polish Academy of Sciences,\\ Aleja Lotnik\'ow 32/46,
    02-668 Warszawa, Poland}
  
  \affiliation{Institute of Theoretical Physics, Faculty of Physics,
University of Warsaw, ul. Pasteura 5, 02-093 Warszawa, Poland
}

\author{Jan Skolimowski}   
  \affiliation{International Research Centre MagTop, Institute of
    Physics, Polish Academy of Sciences,\\ Aleja Lotnik\'ow 32/46,
    02-668 Warszawa, Poland}

\author{Wojciech Brzezicki}   
  \affiliation{International Research Centre MagTop, Institute of
    Physics, Polish Academy of Sciences,\\ Aleja Lotnik\'ow 32/46,
    02-668 Warszawa, Poland}
  
   \affiliation{Institute of Theoretical Physics, Jagiellonian University,\\
       Prof. Stanis\l{}awa \L{}ojasiewicza 11, PL-30348 Krak\'ow, Poland}

\author{Krzysztof Byczuk}   
  \affiliation{Institute of Theoretical Physics, Faculty of Physics,
University of Warsaw, ul. Pasteura 5, 02-093 Warszawa, Poland
}

\author{Marcin M. Wysoki\'nski}   \email{wysokinski@magtop.ifpan.edu.pl}   
  \affiliation{International Research Centre MagTop, Institute of
    Physics, Polish Academy of Sciences,\\ Aleja Lotnik\'ow 32/46,
    02-668 Warszawa, Poland}

 \begin{abstract}
Recently, a quantum anomalous Hall state at odd integer filling in  moir\'e stacked MoTe$_2$/WSe$_2$ was convincingly interpreted as a topological Mott insulator state appearing due to strong interactions in {\it band} basis [P. Mai, J. Zhao, B. E. Feldman, and P. W. Phillips, Nat. Commun. {\bf 14}, 5999 (2023)].
In this work, we aim to analyze the formation of a topological Mott insulator due to interactions in {\it orbital} basis instead, being more natural for systems where interactions originate from the character of $f$ or $d$ orbitals rather than band flatness. For that reason, we study an odd-integer filled Anderson lattice model incorporating odd-parity hybridization between orbitals with different degrees of correlations introduced in the Hatsugai-Kohmoto spirit. 
We demonstrate that a topological Mott insulating state can be realized in a considered model only when weak intra- and inter-orbital correlations involving dispersive states are taken into account. Interestingly, we find that all topological transitions between trivial and  topological Mott insulating phases are not accompanied by a spectral gap closing, consistent  with a phenomenon called {\it first-order topological transition}. Instead, they are signaled by a kink developed in spectral function at one of the time reversal invariant momenta.
We believe that our approach can provide insightful phenomenology of topological Mott insulators in spin-orbit coupled $f$ or $d$ electron systems.
\end{abstract}  

\date{\today}

\maketitle

\section{Introduction}

The interplay between correlations and topology can lead to interesting physical complexity  \cite{Rachel_2018,Amaricci_2015,Maska2018}. However, one of the main challenges that appears in most research problems is the assessment of whether using a well-defined approximation imposed on the many-body interaction  
spoils the reliability of tools utilized for a determination of topological properties that, strictly speaking, apply only to uncorrelated scenarios \cite{atland1997,Chiu13,Chiu14,Sato14,Shio16}.   

In this light, it has been shown that a topological invariant frequently used for correlated systems  constructed from the single-particle Green's function \cite{Zhang_2012,Zhang_2012_2} can lead to false positive information on the topological properties of the system \cite{Yang2019}. Moreover, it has been formally established that such an invariant is not uniquely related to the Hall conductance \cite{arxiv_Zhao,arxiv0}, the ultimate observable related e.g. to the Chern number \cite{Thouless1985}.

In the process of reaching these conclusions, the consideration of a long-range interaction that becomes diagonal in momentum space introduced by Hatsugai and Kohmoto (HK)   \cite{HK}, has turned out to be invaluable. This is because the topological analysis leveraging the single-particle Green's function in a setup with  interactions diagonal in momentum can be compared to the exact calculations of the Berry curvature of the many-body wave function \cite{Yang2019,Wysokinski2023}. Parallel to the important role that  the  HK interaction recently started to play in the exploration of correlated topological systems  \cite{Morimoto2016,Yang2019,Wysokinski2023},  it gained substantial recognition from the perspective of Mott insulator description   
\cite{Nature1, Nature2}. Formerly, it was also explored in the context of statistical spin liquids \cite{Byczuk1994,Byczuk1995} and unconventional superconductivity \cite{Spalek1994}.
 
Lately, the combination of Mott physics due to HK interaction with topological properties  has been explored in Haldane \cite{Mai_2023} and quantum spin Hall models \cite{arxiv2}. This led to the spectacular proposal of the topological Mott insulator at odd integer filling. 
In both works the HK interaction was introduced at the level of band rather than orbital basis, the latter being more natural when constructing Hubbard-type  models \cite{Coe2023}. Nevertheless, Mott physics combined with spontaneous symmetry breaking due to such intraband interaction has been convincingly advocated \cite{arxiv2} as a semiphenomenological rationalization of the quantum anomalous Hall state appearing at oddinteger filling in moir\'e stacked MoTe$_2$/WSe$_2$ \cite{Nature3}. 

In this work, we aim to analyze the topological Mott insulating state from a different perspective. Namely, we introduce HK interactions  at the level of orbitals as in Refs. \cite{Wysokinski2023,Coe2023} but focus on odd-integer fillings; this is more natural for systems where correlations are strong due to the character of $d$ and $f$ orbitals rather than from band flatness as   is the case for moir\'e structures. For that reason, here we analyze a version of the two-dimensional Anderson lattice model \cite{Zhong2022}. 
In its original form the model, incorporating the  Hubbard-type interaction,  displays characteristic properties of many systems whose band structure in the vicinity of the Fermi surface is determined by a hybridization of strongly correlated $f$ or $d$ states with less correlated ligand states. 
For instance it captures a minimal description of
trivial \cite{kondo} and topological \cite{Coleman_2010,Wysokinski_2016} Kondo insulators, metallic ferromagnets \cite{Wysokinski2014R,Wysokinski2015R,Abram2016,Wysokinski2018R,Wysokinski2019}, heavy-fermion Landau liquid \cite{Wysokinski_2015_HF}, unconventional heavy-fermion superconductors \cite{Wysokinski2016} and, most importantly, mixed-valence Mott insulators  at odd-integer fillings \cite{Batista2001,Amaricci2007,Amaricci_2017}. 

Here, for the description of topological Mott insulator state we consider opposite parity orbitals, that for concreteness we assume to be $f$ and $d$ respectively, mixing through the odd parity hybridization emulating the effect of spin-orbit interaction. The correlations incorporated in the HK spirit comprise  $f$ and $d$ intra-orbital interactions (quantified by amplitudes $U_f$ and $U_d$, respectively) as well as $f$-$d$ inter-orbital interaction ($U_{fd}$).
Throughout the whole work we keep the natural hierarchy $U_f>U_{fd}>U_d$ representing a stronger degree of correlations at $f$ orbital than on $d$ orbitals. 

The most important findings of our work are that  ({\it i}) a topological Mott insulator can be stabilized only if intra-$d$-orbital and $f$-$d$ interorbital interactions are present, and ({\it ii})  all topological phase transitions, including transitions between two distinct  topological Mott insulators are not accompanied by a spectral gap closing, consistently with a phenomenon called {\it first-order topological transition}  \cite{Budich2013,Amaricci_2015,Amaricci_2016,Troyer2016,Roy2016,Barbarino2019}. Instead at topological transitions spectral function develops a kink at one of the time reversal invariant momenta (TRIM).  Moreover, we believe that our approach can provide insightful phenomenology of topological Mott insulators in spin-orbit coupled $f$- or $d$-electron systems and thus can pave the way for the search for  topologically nontrivial phases in odd-integer filled $f$ or $d$ orbital systems.

\section{Model}
Our starting point is the two-dimensional Anderson lattice model on a square lattice with the odd-parity hybridization and all-orbital interactions introduced in the Hatsugai-Kohmoto spirit \cite{HK},
\begin{equation}
    \begin{split}
    \mathcal{H}=&\sum_{\g k}\Psi_{\g k}^\dagger\begin{pmatrix}
    (\epsilon_{\g k}-\mu) \mathbb{1}&\mathbb{V}_{\g k}\\
\mathbb{V}_{\g k}&(\epsilon_f-\mu) \mathbb{1}
\end{pmatrix}\Psi_{\g k}+ U_fn^f_{{\g k}\Uparrow } n^f_{{\g k}\Downarrow}\\
&+  U_{fd} (n^d_{{\g k}\uparrow }+n^d_{{\g k}\downarrow })(n^f_{{\g k}\Downarrow }+n^f_{{\g k}\Uparrow })+  U_d n^d_{{\g k}\uparrow } n^d_{{\g k}\downarrow },
  \end{split}
  \label{ham}
  \end{equation}
  where
  $$
\Psi_{\g k}^\dagger =\{d_{{\g k}\uparrow}^\dagger,d_{{\g k}\downarrow}^\dagger,f_{{\g k}\Uparrow}^\dagger,f_{{\g k}\Downarrow}^\dagger\}$$
are fermionic creation operators of particles in a particular orbital and spin states for a given momentum ${\g k}$ (we use the lattice constant $a=1$), 
$$
\mathbb{V}_{\g k} =-2V(\sin k_x\, \sigma_x+\sin k_y\, \sigma_y), 
$$
and
$$
\epsilon_{\g k} =-2t (\cos k_x+\cos k_y). 
$$
Here a dispersion of the conduction band $\epsilon_{\g k}$ is taken in the simplest tight binding form, generated by the nearest neighbor hopping $t$,  which hereafter is taken as an energy unit, i.e. $|t|=1$. Arrows labeling conduction $d$ operators denote spin degrees of freedom. Differently, $f$-operators describe a crystal-field state of a magnetic ion with atomic level $\epsilon_f$ and thus thick arrows labeling these operators stand for a pseudo-spin quantum number \cite{Coleman_2010}. For that reason hybridization operator $\mathbb{V}$ not only comprises a bare hybridization amplitude $V$ but also form factors leading to its odd-parity reflected in specific momentum dependence. In turn, amplitudes $U_d$, $U_{fd}$ and $U_f$ quantify intra-$d$-orbital, inter-$f$-$d$-orbital and intra-$f$-orbital interactions introduced here in Hatsugai-Kohmoto spirit \cite{HK}; though exotic and long-ranged these interactions are representative when it comes to Mott physics as they lead to the same fixed point for Mott transition as more realistic Hubbard-type   interactions \cite{Nature1}. 
Finally, $\sigma_{x,y,z}$ are usual Pauli matrices.

The Hamiltonian (\ref{ham}) does not couple different ${\g k}$ vectors. Moreover, it does not couple sectors describing different numbers of fermions and thus is block diagonal for a given ${\g k}$. Therefore 
we can represent it in the Fock space as
\begin{equation}
   \mathcal{H}= \sum_{{\g k}}\mathcal{H}_{\g k}=\sum_{{\g k}}\sum_n|\hat \alpha^n_{\g k}\rangle  \mathcal{\hat H}^n_{\g k} \langle\hat\alpha^n_{\g k}|
\end{equation}
 where $ \mathcal{\hat H}^n_{\g k}$ describe sectors with an integer number $n\in\{0,1,2,3,4\}$  of particles for a given $\g k$:
 \begin{equation}
    \mathcal{H}^0_{\g k}=0,
\end{equation} 
\begin{equation}
    \begin{split}
        \mathcal{H}_{\g k}^1&=\begin{pmatrix}
        \epsilon_{\g k}\mathbb{1} &\mathbb{V}_{\g k}\\
        \mathbb{V}_{\g k} & \epsilon_f\mathbb{1}
        \end{pmatrix}-\mu \mathbb{1}_4,
    \end{split}
\end{equation}
\begin{equation}
\begin{split}
    &\mathcal{H}_{\g k}^2=(\epsilon_{\g k}+\epsilon_f-2\mu)\mathbb{1}_6\\ 
&+    \begin{pmatrix}
    \begin{bmatrix}
       \epsilon_{\g k}\!-\!\epsilon_f\!+\!U_d&0\\
       0&\epsilon_f\!-\!\epsilon_{\g k}\!+\!U_f\\
    \end{bmatrix}&(\mathbb{1}-\sigma_x)\cdot\mathbb{V}^T &\mathbb{0}\\
   \mathbb{V}^T\cdot(\mathbb{1}-\sigma_x)&U_{fd}\mathbb{1}&\mathbb{0}\\
   \mathbb{0}&\mathbb{0}&U_{fd}\mathbb{1}
    \end{pmatrix},
\end{split}
\end{equation}
\begin{equation}
    \begin{split}
        \mathcal{H}_{\g k}^3&=(\epsilon_{\g k}+\epsilon_f+ 2U_{fd}-3\mu)\mathbb{1}_4\\
        &+\begin{pmatrix}
        (\epsilon_{\g k}+U_d)\mathbb{1} &\mathbb{V}_{\g k}\\
        \mathbb{V}_{\g k} & (\epsilon_f+U_f)\mathbb{1}
        \end{pmatrix},
    \end{split}
\end{equation}
\begin{equation}
     \mathcal{H}^4_{\g k} =2(\epsilon_{\g k}+\epsilon_f)+U_f+U_d+4U_{fd},
\end{equation}
that are spanned by following Fock states for a given ${\g k}$ (where superscript again stands for the number of particles):
 \begin{equation}
     \begin{split}
         |\hat\alpha_{{\g k}}^0\rangle =&\{\ket{  0;\! 0 }_{{\g k}}\}\\
         |\hat\alpha_{{\g k}}^1\rangle =&\{\ket{\uparrow;\!0}_{{\g k}},\ket{\downarrow;\!0}_{{\g k}},\ket{0;\!\Uparrow}_{{\g k}},\ket{0;\!\Downarrow}_{{\g k}}\}\\
         |\hat\alpha_{{\g k}}^2\rangle =&\{\ket{\uparrow \downarrow;\!0}_{{\g k}}, \ket{0;\!\Uparrow \Downarrow}_{{\g k}},\ket{ \downarrow;\!  \Downarrow}_{{\g k}},\ket{ \downarrow;\! \Uparrow }_{{\g k}},\ket{ \uparrow;\! \Uparrow }_{{\g k}},\ket{ \uparrow;\! \Downarrow }_{{\g k}}\}\\
|\hat\alpha_{{\g k}}^3\rangle =&\{\ket{\uparrow \downarrow;\! \Uparrow }_{{\g k}},\ket{\uparrow \downarrow;\! \Downarrow}_{{\g k}},\ket{\uparrow ;\! \Uparrow \Downarrow}_{{\g k}},\ket{ \downarrow;\! \Uparrow \Downarrow}_{{\g k}}\},\\
|\hat\alpha_{{\g k}}^4\rangle =&\{ \ket{\uparrow \downarrow;\! \Uparrow \Downarrow}_{{\g k}}\}.
 \end{split}
 \end{equation}
 The first (second) sector inside a Dirac ket symbol refers to the $d$ ($f$) orbital. 
Note that despite the presence of the many-body interaction, the momentum remains a good quantum number and thus our model is exactly solvable. 

In this work, in addition to the consideration of the exact eigenfunctions and eigenenergies ($E_{{\g k},m}$) we are analyzing the zero-temperature spectral function that in the Lehmann representation reads
\begin{equation}
\begin{split}
    \mathcal{A}({\g k}, \omega)=\sum_{\alpha }\sum_{m,m'}g_{mm'}\langle m|\alpha_{\g k }&|m'\rangle\langle m'|\alpha^\dagger_{\g k}|m\rangle \\ &\times\delta(\omega+E_m-E_{m'})
\end{split},
    \end{equation}
where the operators are $\alpha_{\g k} \in\{d_{{\g k} \uparrow},d_{{\g k} \downarrow},f_{{\g k} \Uparrow} ,f_{{\g k} \Downarrow} \}$ and
\begin{equation}
    g_{mm'}=\lim_{\beta\to\infty} \frac{{\rm e}^{-\beta E_m}+{\rm e}^{-\beta E_{m'}}}{\mathcal{Z}},
\end{equation}
that for a two-fold degenerate groundstate (at given $\g k$) takes on values $\{0,1/2,1\}$. $\mathcal{Z}$ is a partition function.

\section{The symmetries of the model \label{sec:sym}}

To define symmetries of the many-body Hamiltonian it is convenient to first define two basic operations on fermions. The first one is the fermion-interchange operator:
%
\begin{equation}
    C_{ij} =  \mathbb{1} - (\Psi^{\dagger}_{i}-\Psi^{\dagger}_{j}) (\Psi_{i}-\Psi_{j}),
\end{equation}
with $\Psi^{\dagger}$ being a vector of four fermion operators defined under Eq. (1). Here we omit the momentum $\g k$ for simplicity. The second operation is the fermion-phase operator:
\begin{equation}
    G_{i}(\phi) =  e^{i\phi}\Psi^{\dagger}_{i}\Psi_{i} + \Psi_{i}\Psi^{\dagger}_{i}.
\end{equation}
Their action on fermion operators is:
\begin{eqnarray}
C_{ij}\Psi_k C_{ij} &=& \delta_{jk}\Psi_i+\delta_{ik}\Psi_j+\varepsilon^2_{ijk} \Psi_k\nonumber\\
G_{i}(\phi)\Psi_j G_{i}(-\phi) &=& e^{-i\phi}\delta_{ij}\Psi_j+\varepsilon^2_{ij} \Psi_j,
\end{eqnarray}
having that $C_{ij}C_{ij}=G_{i}(\phi)G_{i}(-\phi)=\mathbb{1}$. Above, $\varepsilon$ is the Levi-Civita symbol and $\delta$ is the Kronecker delta.
Now we define the time-reversal operator as:
\begin{eqnarray}
{\cal T}={\cal K}\,C_{12}G_1(\tfrac{\pi}{2})G_2(-\tfrac{\pi}{2})C_{34}G_3(\tfrac{\pi}{2})G_4(-\tfrac{\pi}{2}),
\end{eqnarray}
where ${\cal K}$ is complex conjugation. The action of $\cal T$ on $d$ and $f$ fermions is given by:
\begin{eqnarray}
{\cal T} d_{\uparrow} {\cal T}^{-1} &=& i d_{\downarrow},\quad {\cal T} d_{\downarrow} {\cal T}^{-1} = -i d_{\uparrow}\nonumber\\
{\cal T} f_{\Uparrow} {\cal T}^{-1} &=& i f_{\Downarrow},\quad {\cal T} f_{\Downarrow} {\cal T}^{-1} = -i f_{\Uparrow}.
\end{eqnarray}
Here without loss of generality we assume that all fermion operators are defined as real.
Having the above expressions it is easy to find that $\mathcal{T}$ satisfies a typical relation with 
the Hamiltonian. For fixed momentum ${\g k}$ Hamiltonian $\mathcal{H}_{\g k}$ can be 
written as
\begin{equation}
\mathcal{H}_{\g k}=\Psi^\dagger H_{\g k}\Psi+ \mathcal{H}^{\rm int}
\end{equation}
where we put all the fourth order terms in fermion operators in $\mathcal{H}^{\rm int}$.
Now we notice that the action of the time-reversal operator on $\mathcal{H}_{\g k}$ is given by
\begin{equation}
{\cal T} {\cal H}_{{\g k}} {\cal T}^{-1} = \Psi^\dagger H_{\g -k}\Psi + \mathcal{H}^{\rm int} = {\cal H}_{{\g -k}},
\end{equation}
which reduces to the well-know time-reversal symmetry (TRS) on a non-interacting system when $\mathcal{H}^{\rm int}_{\g k}=0$. Note that fermions $\Psi$ carry no {\it explicit} dependence on momentum $\g k$ so for simplicity it can be omitted in the above formula.  

Similarly to the Hamiltonian itself the TRS operator can be written in a block diagonal form in the basis with fixed number of particles. Hence, we get the following blocks:
\begin{eqnarray}
\mathcal{T}^0&=&\mathcal{T}^4=\mathcal{K},\nonumber\\
\mathcal{T}^1&=&-\mathcal{T}^3=-\mathcal{K}\,\mathbb{1}_2\otimes\sigma_y,\nonumber\\
\mathcal{T}^2&=&-\mathcal{K}\,\mathbb{1}_2\oplus\sigma_x\oplus(-\sigma_x)
\end{eqnarray} 
satisfying: 
\begin{equation}
\mathcal{T}^n\mathcal{H}^n_{\g k}(\mathcal{T}^n)^{-1} = \mathcal{H}^n_{-\g k},
\end{equation}
for $n=0,1,\dots,4$ and $\mathcal{H}^n_{\g k}$ being matrix blocks defined in the previous 
section. Quite surprisingly, we find  that for different particle numbers $n$ we can get 
different symmetry class of the $\mathcal{H}^n_{\g k}$ block. 
More specifically, we get $(\mathcal{T}^n)^2=-1$ for every odd $n$ that yields AII 
symmetry classes of the $\mathcal{H}^{1,3}_{\g k}$ blocks and $(\mathcal{T}^n)^2=1$ for all 
even $n$ which gives AI symmetry class of the $\mathcal{H}^{0,2,4}_{\g k}$ blocks.

Another important symmetry of the model is inversion. It can be defined using fermion-phase operators as:
\begin{equation}
P = G_3(\pi)G_4(\pi),
\end{equation}
yielding
\begin{eqnarray}
P d_{\uparrow}P^{-1}&=&d_{\uparrow},\quad P d_{\downarrow}P^{-1}=d_{\downarrow},\nonumber\\
P f_{\Uparrow}P^{-1}&=&-f_{\Uparrow},\quad P f_{\Downarrow}P^{-1}=- f_{\Downarrow},
\end{eqnarray} 
and having unitary relation with the Hamiltonian:
\begin{equation}
P {\cal H}_{{\g k}} P^{-1} ={\cal H}_{{\g -k}}.
\end{equation}
Inversion preserves the number of particles so it can be decomposed into  blocks analogous to   the Hamiltonian and TRS operator:
\begin{eqnarray}
P^0&=&P^4=1,\nonumber\\
P^1&=&-P^3=-\sigma_z\otimes\mathbb{1}_2,\nonumber\\
P^2&=&\mathbb{1}_2\oplus(-\mathbb{1}_2)\oplus(-\mathbb{1}_2).
\end{eqnarray} 
Due to the presence of TRS such that $(\mathcal{T}^n)^2=-1$ in the odd-$n$ blocks, the inversion symmetry will make every state twice degenerate at every $\g k$-points for $n\in\{1,3\}$.  

Finally, the model exhibits additional unitary symmetry that allows one to decompose each $\mathcal{H}^n_{\g k}$ block into two subblocks. Again, it can be defined using fermion-phase operators as 
\begin{equation}
\Sigma = G_2(\pi)G_3(\pi),
\end{equation}
yielding
\begin{eqnarray}
\Sigma d_{\uparrow}\Sigma^{-1}&=&d_{\uparrow},\quad\,\,\,\,\, \Sigma d_{\downarrow}\Sigma^{-1}=- d_{\downarrow}\nonumber\\
\Sigma f_{\Uparrow}\Sigma^{-1}&=&- f_{\Uparrow},\quad \Sigma f_{\Downarrow}\Sigma^{-1}=f_{\Downarrow}
\end{eqnarray} 
and leading to the commutation relation with the Hamiltonian:
$[\Sigma, {\cal H}_{{\g k}}] = 0$. In the fixed-$n$ subblocks this symmetry takes the form of
\begin{eqnarray}
\Sigma^0&=&\Sigma^4=1,\nonumber\\
\Sigma^1&=&\Sigma^3=\sigma_z\otimes\sigma_z,\nonumber\\
\Sigma^2&=&(-\mathbb{1}_2)\oplus(-\mathbb{1}_2)\oplus\mathbb{1}_2.
\end{eqnarray}

\section{Topological Mott states at odd-integer fillings: general considerations}

To realize a topological Mott insulating state at odd-integer filling within the considered two-dimensional model,  two elements  need  to be simultaneously satisfied. Namely, the system, due to many-body interactions, must open a Mott insulating gap and the range of considered parameters  must provide a nontrivial topology of the ground state.  

Meeting the first condition is signaled by the opening of the spectral gap at odd integer filling when interactions are sufficiently strong. Although we determine the metal-insulator transition line with respect to model parameters numerically, for future discussion it is also instructive to formulate an analytical condition for opening a Mott gap. Namely, the realization of an insulating state at odd integer filling means that for all momenta ${\g k}$, the ground state of  $\mathcal{H}_{\g k}$ Hamiltonian lies in the sector spanned solely by $|\hat\alpha^1_{\g k}\rangle$, for filling 1 electron per lattice site, and $|\hat\alpha^3_{\g k}\rangle$ , for filling 3, and there is a non-zero gap to the first excited state that changes the number of electrons ensuring opening of the spectral Mott gap. Practically, if by $E^{(n)}_{{\g k}}$ we denote the lowest eigenvalue in the subsector with $n$ electrons, we can write the above condition as
\begin{equation}
    \forall_{{\g k},n\neq 1}\,\,\, E^{(1)}_{\g k}< E^{(n)}_{{\g k}} \label{c1}
\end{equation}
for filling 1, and
\begin{equation}
    \forall_{{\g k}, n\neq 3}\,\,\, E^{(3)}_{\g k}< E^{(n)}_{{\g k}} \label{c3}
\end{equation}
for filling 3.

\begin{figure}
    \centering
    \includegraphics[width=0.4  \textwidth]{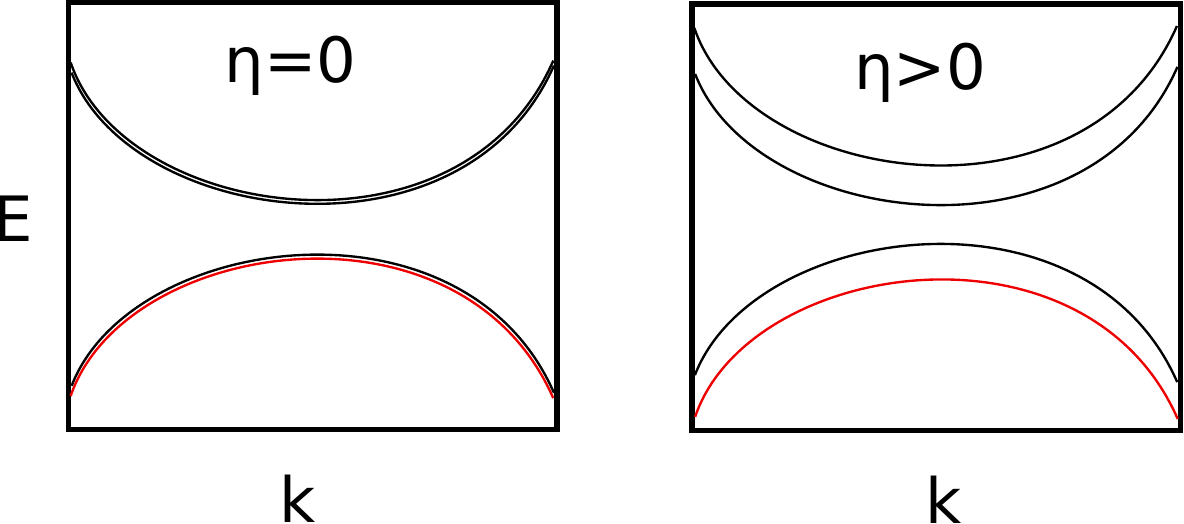}
    \caption{Schematic demonstration that arbitrary small $\eta>0$ in $\mathcal{H}^{1(3)}_{\g k}$ [Eq. \eqref{eta}] yields a gapped and unique groundstate as a consequence of eigenstates of $\mathcal{H}^{1(3)}_{\g k}$ being many-body Fock states instead of single particle states. Red color denotes states that can be considered as filled in terms of a single-particle system analogy.}
    \label{fig_s}
\end{figure}

On the other hand, the determination of topological properties of the Mott insulating states at odd integer filling is more convoluted because of the high degeneracy of the ground state. Although introducing HK interaction in orbital basis removes the degeneracy at half filling \cite{Wysokinski2023, Coe2023} it does not help in the case of the odd-integer filled state, where two-fold Kramers degeneracy is present at each $\g k$ due to the combined $\mathcal{T}$ and $P$ symmetry. 
However, we can use symmetry $\Sigma$ to 
split the Hamiltonians $\mathcal{H}_{\g k}^{1(3)}$
into subblocks. We find that:
\begin{equation}
     \begin{split}
     \mathcal{H}_{\g k}^{1(3)}=\begin{pmatrix}
         \mathcal{H}_{\g k+}^{1(3)}&0\\ 
         0&\mathcal{H}_{\g k-}^{1(3)}
     \end{pmatrix}
     \end{split}
 \end{equation}
 where 
\begin{equation}
\begin{split}
    \mathcal{  H}_{{\g k}\pm}^1=& \begin{pmatrix}
        \epsilon_{\g k}-\mu& 2V(\pm i\sin k_y\!-\!\sin k_x)\\
        2V(\mp i\sin k_y\!-\!\sin k_x)&\epsilon_f-\mu
    \end{pmatrix},\\
    \mathcal{  H}^3_{{\g k}\pm}=& \begin{pmatrix}
        \epsilon_{\g k}-\mu+U_d& 2V(\pm i\sin k_y\!-\!\sin k_x)\\
        2V(\mp i\sin k_y\!-\!\sin k_x)&\epsilon_f-\mu+U_f
    \end{pmatrix}\\
    &+(\epsilon_{\g k}+\epsilon_f+2U_{fd}-2\mu) \mathbb{1}.
\end{split} 
\end{equation}
Since $\mathcal{T}$ and $P$ anticommute in the 
odd-particle blocks the subblocks 
$\mathcal{H}_{\g k+}^{n}$ and $\mathcal{H}_{\g 
k-}^{n}$ are time-reversal partners and each of 
them taken separately breaks TRS. This allows for 
nonzero Chern numbers of the eigenstates of the  $\mathcal{H}_{\g k\pm}^{1(3)}$ Hamiltonian that can be easily calculated from the decomposition
\begin{equation}
\mathcal{H}_{\g k\pm}^{1(3)} = {\g b}_{\g k}\cdot \boldsymbol{\sigma}  + a_{\g k}\mathbb{1}.
\end{equation}
We can now define Berry curvature as
\begin{equation}
\Omega_{\g k\pm}^{1(3)} = \frac{1}{2|{\g b}|^3} {\g b}\cdot\big(\partial_{k_x}{\g b}\times\partial_{k_y}{\g b}\big),
\end{equation}
and the Chern number of the bottom band of $\mathcal{H}_{\g k\pm}^{1(3)}$ is given by
\begin{equation}
C_{\pm}^{1(3)} = \frac{1}{ 2\pi} \int_{{\g k}}   \!\!d^2 {\g k}\,\,\Omega_{\g k\pm}^{1(3)}.
\label{chern}
\end{equation}
The ground-state manifold consists of the degenerate states
\begin{equation}
\ket{\psi^{1(3)}_s} = \bigotimes_{{\g k}\in BZ}\ket{\phi^{1(3)}_{s(\g k)}(\g k)},
\label{psis}
\end{equation}
where $\ket{\phi^{1(3)}_{s(\g k)}(\g k)}$ is the lowest energy eigenstate of $\mathcal{H}_{\g k}^{1(3)}$ living in the $\Sigma = \pm 1$ subspace and $s: BZ\to \{-1,1\}$ is any function. 
Obviously the $\Sigma$-resolved Chern number described above can be defined only in the special cases of $s(\g k)\equiv 1$ or $s(\g k)\equiv -1$ because only in this case is the ket on the right-hand side of Eq. (\ref{psis}) a continuous function of $\g k$.
For all other cases we cannot define the Chern number but it still carries   useful information about the system. Having well-defined $C_{+}^{1(3)}$, we know that the system has non-zero excitation gap between the highly degenerate ground state and the lowest excited state. We also know that if we have two ground states differing by the value of $C_{+}^{1(3)}$ then there must be a closing of this gap when we go from one to the other. Note that the topological nature of the ground-state manifold can be easily revealed by a small perturbation added to the Hamiltonian, 
\begin{equation}
     \mathcal{H}_{\g k}^{1(3)}\to \mathcal{H}_{\g k}^{1(3)}- \eta\sigma_z\otimes\mathbb{1}_2. \label{eta}
\end{equation}
This acts as a chemical potential with opposite sign for the two  subblocks, $\mathcal{H}_{{\g k}\pm}^{1(3)}\to \mathcal{H}_{{\g k}\pm}^{1(3)} \mp\eta\mathbb{1}_2$, breaking the TRS of $\mathcal{H}_{\g k}^{1(3)}$ and lifting the degeneracy of the lowest-energy eigenstate. Now the ground state is unique and can have a Chern number $C_{+}^{1(3)}$ for $\eta>0$ or $C_{-}^{1(3)}$ for $\eta<0$ in strict analogy to the ground state of the ionic Rashba model with HK interaction considered in Ref. \cite{Wysokinski2023}. Note that parameter $\eta$ can be arbitrarily small to trigger the Chern insulating phase because it does not have to open a direct gap. In contrast to the non-interacting case an indirect gap between the first and second eigenstate of $\mathcal{H}_{\g k}^{1(3)}$ is enough to yield a gapped ground state; see Fig. \ref{fig_s}.

The $\Sigma$-resolved Chern number is analogous to the spin-resolved Chern number that can be defined in a 2D topological insulator (TI) if the spin-orbit coupling is zero. However, the question remains what happens if it is not. For a 2D TI we are able to define a $\mathbb{Z}_2$ invariant protected by TRS. Here it is also possible, to some extent, but with one important difference. In a standard TI we always have even filling because, due to the Kramer degeneracy at TRIMs it is not possible to have a gap otherwise. In the present case the filling is effectively fixed to be 1, for both   $\mathcal{H}_{\g k}^{1}$ and  $\mathcal{H}_{\g k}^{3}$, because for every $\g k$-point we have to choose one lowest-energy many-body state. Therefore 
it seems there is no way to have a gap. Indeed, if we add a small perturbation to our Hamiltonian, namely
\begin{equation}
     \mathcal{H}_{\g k}^{1(3)}\to \mathcal{H}_{\g k}^{1(3)}+ \xi(\sin k_x + \sin k_y)\sigma_y\otimes\mathbb{1}_2.
\end{equation}
we will find that $\Sigma$ symmetry is broken and $\mathcal{H}_{\g k}^{1(3)}$ cannot be decomposed into subblocks having Chern numbers. Apart from this the energy spectrum still looks the same because TRS and inversion symmetry are preserved. The ground state will be again an exponentially degenerate multiplet of states given by Eq. (\ref{psis}) separated by the gap from the lowest excited state. This energy gap will be again equivalent to the energy gap of $\mathcal{H}_{\g k}^{1(3)}$ between its second and third eigenvalues. Now, since $\mathcal{H}_{\g k}^{1(3)}$ is an AII-symmetry-class matrix being a function of two continuous and periodic parameters $k_x$ and $k_y$ we can ask about its $\mathbb{Z}_2$ topological invariant related to this energy gap. This is equivalent to treating $\mathcal{H}_{\g k}^{1(3)}$ as if it was a free-fermion Hamiltonian at half-filling. 
Having the inversion symmetry we can follow Refs.~[\onlinecite{Kane2005}]~and~[\onlinecite{Fu2007}], and calculate $\mathbb{Z}_2$ invariant for $\mathcal{H}^{1(3)}$.  To do this we need to inspect $\mathcal H^{1(3)}$ Hamiltonians at TRIMs ${\g k}={\g k^*}$, which for the square lattice are ${\g k^*}\in\{\Gamma,X,Y,M\}$. They read:
\begin{equation}
   \mathcal{H}^1_{\g k^*}= \frac{1}{2}\big[(\epsilon_{\g k^*}+\epsilon_f)\mathbb{1}_4+(\epsilon_{\g k^*}-\epsilon_f)\sigma_z\otimes\mathbb{1}_2\big]
\end{equation}
for   filling 1, and
\begin{equation}
   \mathcal{\tilde H}^3_{\g k*}= \frac{1}{2}\big[(\epsilon_{\g k*}+\epsilon_f+U_f+U_d)\mathbb{1}_4+(\epsilon_{\g k*}-\epsilon_f+U_d-U_f)\sigma_z\otimes\mathbb{1}_2\big]
\end{equation}
for   filling 3. Now the $\mathbb{Z}_2$ topological invariant $\nu$ can be readily obtained from eigenvalues of the inversion symmetry operator $P^{1(3)}=\mp\sigma_z\otimes\mathbb{1}_2$ as
\begin{equation}
    (-1)^\nu=\prod_{{\g k}^*\in\{\Gamma,M,X,Y\} }\delta_{\g k^*} \label{nu},
\end{equation}
where for  filling 1 
\begin{equation}
    \delta_{\g k^*} ={\rm sgn}(\epsilon_{{\g k}^*}-\epsilon_f),
\end{equation}
and for   filling 3
\begin{equation}    
\delta_{\g k^*} ={\rm sgn}(\epsilon_{{\g k}^*}-\epsilon_f+U_d-U_f).
\label{to}
\end{equation}
These considerations allows us to analytically  determine boundaries between topological and trivial insulating 
phases of the model.
In result, in addition to the metallic and topologically trivial Mott insulating (MI) states we distinguish two distinct topological Mott insulator (TMI) phases, TMI(M) with parities of the auxiliary model $\delta_{\g k^*}=\{-1,1,-1,-1\}$ and TMI($\Gamma$) with parities $\delta_{\g k^*}=\{-1,1,1,1\}$.
Note that the distinction between phases TMI($\Gamma$) and TMI(M) follows from the presence of translation and inversion symmetry on top of TRS. The TRS alone would give only one nontrivial phase.

\section{Absence of topological Mott state for $U_d=U_{fd}=0$}

In the absence of intra-$d$-orbital and inter-$f$-$d$-orbital  interactions ($U_d=U_{fd}=0$), although Mott insulating phases are generically realized when $U_f$ is sufficiently strong, they are always topologically trivial, as the following analytical considerations prove.

As we noted in the previous section, opening of the Mott gap is equivalent to satisfying conditions \eqref{c1} and \eqref{c3} when the filling of the system is  1 or 3, respectively. These conditions are formulated for all ${\g k}$ and, in particular, need to be satisfied also at the TRIM. 
At the TRIM the hybridization vanishes (i.e., $\mathbb{V}_{\g k}=\mathbb{0}$) and thus eigenvalues   of the Hamiltonians  $\mathcal{H}_{{\g k}^*}^n$ are defined simply by their diagonal matrix elements. Explicitly, the lowest energy states $E_{{\g k}^*}^{(n)}$ of  $\mathcal{H}^n_{\g k^*}$ read 
\begin{equation}
 \begin{split}
     E_{{\g k}^*}^{(0)}&=0,\\
     E_{{\g k}^*}^{(1)}&=-\mu +{\rm min}(\epsilon_{{\g k}^*},\epsilon_f),\\
     E_{{\g k}^*}^{(2)}&=\epsilon_{{\g k}^*}+\epsilon_f-2\mu+{\rm min}(\epsilon_{{\g k}^*}-\epsilon_f,\epsilon_f-\epsilon_{{\g k}^*}+U_f,0),\\
     E_{{\g k}^*}^{(3)}&=2\epsilon_{{\g k}^*}+\epsilon_f-3\mu+{\rm min} (\epsilon_f-\epsilon_{{\g k}^*}+U_f,0)\\
     E_{{\g k}^*}^{(4)}&= 2(\epsilon_{{\g k}^*}+\epsilon_f)-4\mu+U_f.
 \end{split}   
\end{equation}
After some algebra, the  condition for realizing  the Mott insulating state at   filling 1 reduces
to
\begin{equation}
   \forall_{\g k^*}\,\, \epsilon_{{\g k}^*}>\epsilon_f.
    \label{mott_cond1}
\end{equation}
On the other hand when it comes to topology, analysis of the $\mathbb{Z}_2$ invariant for $H^1_{\g k}$ sets the bound for the topologically nontrivial system to be 
\begin{equation}
    \exists_{{\g k}^*}\,\,\, \epsilon_f>\epsilon_{{\g k}^*}
\end{equation}
The two findings contradict   each other, proving that if $U_d=U_{fd}=0$ a topological Mott insulating state does not exist at the filling 1 in the considered model.

\begin{figure}
    \centering
    \includegraphics[width=0.45\textwidth]{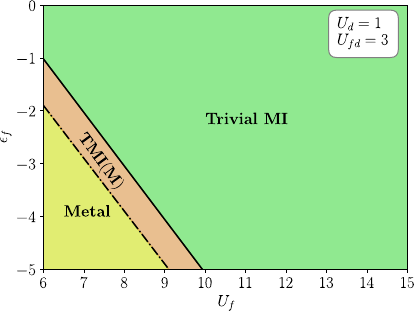}
    \caption{Phase diagram on the $\epsilon_f$-$U_f$ plane for   filling 3. Dashed-dot and solid lines mark the metal/Mott insulator and the topological transitions, respectively.  Non-zero $U_d$ and $U_{fd}$ are essential for the stabilization of the topological Mott insulator phase in a region near the metal/insulator transition. Here $U_d=1$, $U_{fd}=3$, and $V=0.5$.}
    \label{fig0}
\end{figure}

For the system with   filling 3, we focus only on the M point, where the energy $\epsilon_{\g k}$ becomes maximal  ($\epsilon_M\equiv\epsilon_{\g k=\{\pi,\pi\}}=4$).
Then the condition for stabilizing the Mott insulator  can be simplified (under the typical assumption that $\epsilon_f<0$) to 
\begin{equation}
    U_f+\epsilon_f>\mu \,\,\,\,\,\,\,\,\,\, \& \,\,\,\,\,\,\,\,\,\, \mu>\epsilon_M.
    \label{mott_cond}
\end{equation}
In turn, the $\mathbb{Z}_2$ invariant for $\mathcal{\tilde H}^3_{\g k}$ sets the bound for the topologically nontrivial system to be 
\begin{equation}
    U_f+\epsilon_f<\epsilon_{M}.
\end{equation}
Again the two  findings contradict each other and thus we prove that if $U_d=U_{fd}=0$ and $\epsilon_f<0$ a topological Mott insulating state does not exist also at    filling 3.

\begin{figure*} 
    \centering
    \includegraphics[width=1\textwidth]{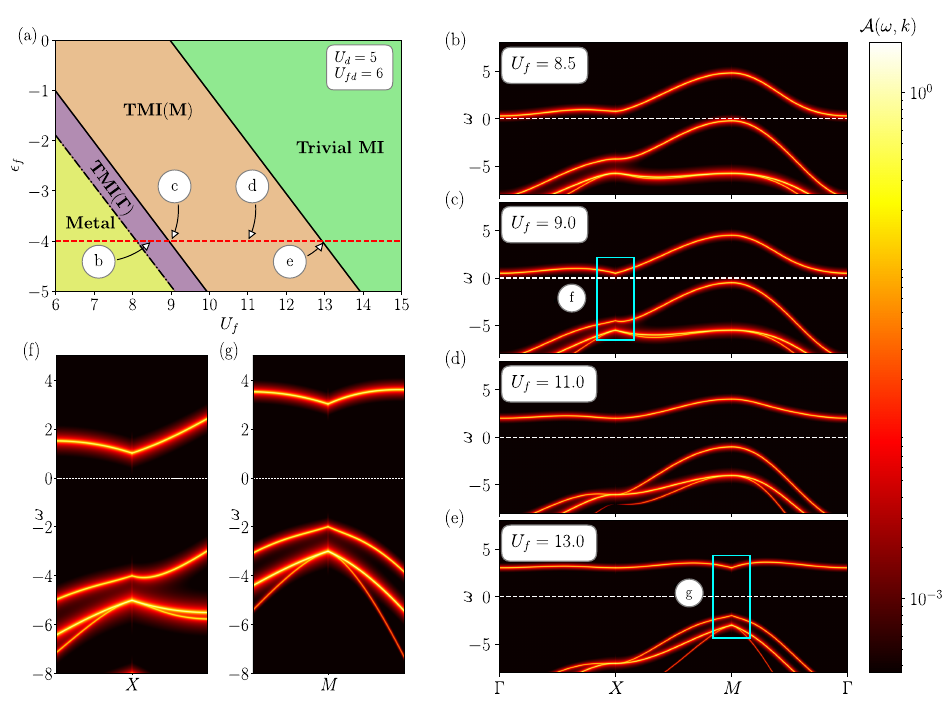}
    \caption{(a)  Phase diagram on the $\epsilon_f$-$U_f$ plane for the filling 3. Dashed-dot and solid lines mark the metal/Mott insulator and the topological transitions, respectively. Relatively large $U_d$ and $U_{fd}$ leads to significantly enlarged (cf., Fig.\ref{fig0})  stability region of the topologically non-trivial states and unveiling of  TMI($\Gamma$) state. (b-e) Spectral functions (color scale  in arbitrary units) along the high symmetry lines for the square lattice in the vicinity of the Fermi level for selected values of the intra-$f$-orbital interaction strengths $U_f$, providing different topological phases and the topological phase transitions along the red dashed-line in the phase diagram (a) for a fixed $\epsilon_f=-4$: (b)  topological Mott insulator TMI($\Gamma$), (c) topological transition TMI($\Gamma$)/TMI(M), (d) topological Mott insulator TMI(M), (e) topological transition TMI(M)/trivial Mott insulator. Instead of the spectral gap closing at the topological phase transitions (c, e) the system develops  kinks  at selected TRIMs, which are zoomed in the panels (f) and (g) with the range of parameters corresponding to marked blue rectangles in panels (c) and (e) respectively. Here $U_d=5$, $U_{fd}=6$, and $V=0.5$.
    }
    \label{fig1}
\end{figure*}

\section{Results}

In this section we present our results concerning the appearance of the topological Mott insulator (TMI) in the considered model. Here we focus solely on the filling 3, while we will briefly discuss the filling 1 case in the Appendix.

In the previous section we   provided the {\it no-go} theorem proving the absence of TMI  if $U_d=U_{fd}=0$. For that reason here we assume that both $U_d$ and $U_{fd}$ amplitudes are non-zero. From Eqs. \eqref{nu} and  \eqref{to} it is clear that an increase of $U_d$ amplitude shifts borders of a potentially topological state towards higher values of $U_f$, thus possibly yielding stable TMI. This however is not the case when $U_{fd}$ remains 0 as the metal/Mott insulator transition is shifted along to higher values of $U_f$ and effectively no TMI is present. The situation changes for the non-zero $U_{fd}$. Namely, we found that, for a considered value of $V=0.5$,  only if both non-zero $U_{d}$ and $U_{fd}$  satisfying proper hierarchy ($U_d<U_{fd}<U_f$) are taken into account, does TMI
appear  stable near the metal/Mott insulator transition. In Fig.~\ref{fig0} we show the phase diagram for the parameters reflecting relative scales expected in the $f$-$d$  orbital system, i.e. for rather small $U_d=1$ and a bit larger $U_{fd}=3$ when compared to $U_f$, which at the phase diagram ranges from 6. As a result, our phenomenological approach to the spin-orbit coupled $f$-orbital Mott insulators indicates that topological Mott insulators should be sought in systems being on the verge of the metal/insulator transition and with ligands states in the vicinity of the Fermi level derived from $d$ orbitals, to justify non-negligible correlations among them.

In turn, Fig.~\ref{fig1}a shows the phase diagram where we increase the $U_d$ and $U_{fd}$ values, still formally satisfying proper hierarchy $U_d<U_{fd}<U_f$. In this case, the stability region of the topological phase significantly increases, which allowed unveiling the  TMI($\Gamma$) phase in addition to TMI(M)   realized already for smaller values of $U_d$ and $U_{fd}$ (cf. Fig. \ref{fig0}). This provides an opportunity to analyze the character of the topological phase transitions not only between topological and trivial phases \cite{Wysokinski2023} but also between two different topological phases.

In Figs.~\ref{fig1}b-e we plot spectral functions in the vicinity of the Fermi level for TMI($\Gamma$) and TMI(M)  phases and at all topological phase transitions along a path with fixed $\epsilon_f=-4$, marked with red dashed line in Fig.~\ref{fig1}a. We found that both topological transitions, between TMI($\Gamma$) and TMI(M) as well as between TMI(M) and trivial MI are not associated with the spectral gap closing, consistently with a phenomenon called {\it first-order topological transition} \cite{Budich2013,Amaricci_2015,Amaricci_2016,Troyer2016,Roy2016,Barbarino2019}.
It is a result  analogous to the one obtained in Ref.~[\onlinecite{Wysokinski2023}] where the phase transition between quantum anomalous Hall insulator and the trivial phases has been found to take place without the spectral gap closing. 

\begin{figure}
    \centering
    \includegraphics[width=0.5\textwidth]{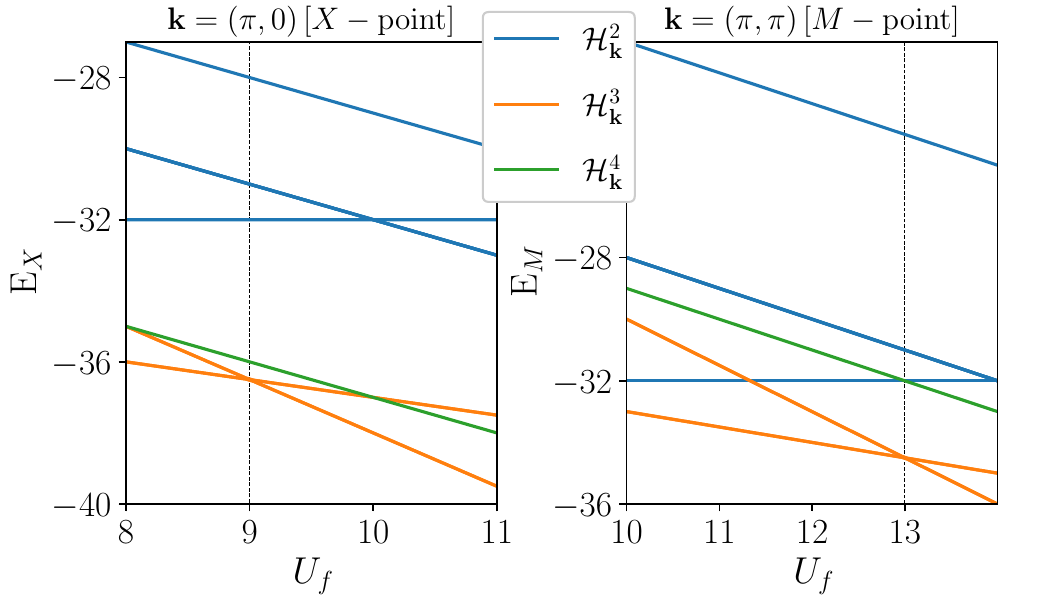}
    \caption{Eigenenergies of the Hamiltonian $\mathcal{H}$ (filling 3) at time reversal invariant momenta (a) X  and  (b) M versus changing  the intra-$f$-orbital interaction $U_f$ across the topological phase transition (vertical dashed line) between (a) TMI($\Gamma$) and TMI(M) and (b) TMI(M) and trivial MI. With separate colors, we denote eigenenergies belonging to the subsectors with a particular number of fermions. Here $U_d=5$, $U_{fd}=6$, and $V=0.5$.   }
    \label{fig2}
\end{figure}

Instead of the spectral gap closing, here both topological phase transitions are signaled by appearance of the kink as  exposed  in Figs.~\ref{fig1}~f~and~g showing the zoomed spectral function.
This is slightly different from the case of the quantum anomalous Hall/trivial insulator transition analyzed in Ref.~[\onlinecite{Wysokinski2023}], where the spectral function at the transition changed discontinuously.

For a better understanding of the described feature in Fig.~\ref{fig2} we plot eigenenergies at the TRIM X  (panel a) and M (panel b) points with an increasing $U_f$ across the topological transitions between TMI($\Gamma$) and TMI(M) (panel a) and TMI(M) and trivial MI (panel b). Both topological phase transitions take place in a usual, continuous manner, namely, when the eigenenergies corresponding to the ground state $|\psi_1\rangle$  and the lowest excited state $|\psi_2\rangle$  cross and thus become degenerate. The difference between such behavior and the one observed for uncorrelated systems is that here both states belong to the same subsector  describing  a fixed particle number. 

\section{Filling 2}
Although this work is devoted to Mott insulating phases at odd-integer filling, it is fair to comment on the half filling case, as the model in the version with the Hubbard term is mostly recognized to realize a topological Kondo insulating phase \cite{Werner2013,Yoshida2013}, at the mean-field level being just a $Z_2$ topological insulator with correlation-renormalized parameters \cite{Coleman_2010, Wysokinski_2016}. As settled in Sec. \ref{sec:sym} for filling with two electron per each $\g k$-point we obtain Hamiltonian  $\mathcal{H}_{\g k}^2$ with TRS given by $\mathcal{T}^2=-\mathcal{K}\,\mathbb{1}_2\oplus\sigma_x\oplus(-\sigma_x)$ and inversion symmetry by $P^2=\mathbb{1}_2\oplus(-\mathbb{1}_2)\oplus(-\mathbb{1}_2)$. This means that $\mathcal{H}_{\g k}^2$ is an AI-symmetry-class matrix dependent on two periodic parameters $k_x$ and $k_y$. Therefore, from the general symmetry classification, we know that it must be topologically trivial. Here it is even possible to show more. Since we have both TRS and inversion symmetry we can combine both to get the so-called $PT$ symmetry. In the present case this $PT$ symmetry squares to $+1$, thus it is possible to use its eigenbasis to put the Hamiltonian $\mathcal{H}_{\g k}^2$ in a purely real form. Indeed we find that under a unitary transformation,
\begin{equation}
\mathcal{U}=(i\mathbb{1}_2)\oplus\frac{1}{\sqrt{2}} 
\begin{pmatrix}
i & 1 \\
-i & 1
\end{pmatrix}
\oplus\mathbb{1}_2
\end{equation}
the Hamiltonian becomes
\begin{equation}
\begin{split}
&\tilde{\mathcal{H}}_{\g k}^2=\mathcal{U}^{\dagger}\mathcal{H}_{\g k}^2\mathcal{U}=(\epsilon_{\g k}+\epsilon_f-2\mu)\mathbb{1}_6\\ 
&+    \begin{pmatrix}
    \begin{bmatrix}
       \epsilon_{\g k}\!-\!\epsilon_f\!+\!U_d&0\\
       0&\epsilon_f\!-\!\epsilon_{\g k}\!+\!U_f\\
    \end{bmatrix}&
   \tilde{\mathbb{V}}_{\g k}&\mathbb{0}\\
   \tilde{\mathbb{V}}^{T}_{\g k}&U_{fd}\mathbb{1}&\mathbb{0}\\
   \mathbb{0}&\mathbb{0}&U_{fd}\mathbb{1}
    \end{pmatrix},
\end{split}
\end{equation}
with 
\begin{equation}
\tilde{\mathbb{V}}_{\g k}=2\sqrt{2}V
\begin{pmatrix}
      \sin k_x & -\sin k_y\\
      -\sin k_x & \sin k_y
   \end{pmatrix}.
\end{equation}
From the reality of $\tilde{\mathcal{H}}$ we have that all its eigenstates have to be real. This means that the Berry curvature of every eigenstate will vansish at every $\g k$ - point, excluding any Chern or valley-Chern number. However, such structure of the Hamiltonian supports point gap closing points at an arbitrary $\g k$ - point, protected by a quantized Zak phase in a closed loop surrounding it.

\section{Summary and discussion}

In the present work, we have studied a topological Mott insulator realized by the two-dimensional Anderson lattice model describing a lattice of strongly correlated, localized  $f$ orbital states mixing with the odd-parity hybridization to weakly correlated ligand $d$ states, where all intra and inter-orbital interactions are incorporated in the Hatsugai-Kohmoto spirit \cite{HK}. 
We have shown that the topological Mott insulating phases are realized only if both the intra $f$-orbital correlations and the intra-$d$-orbital and inter-$f$-$d$-orbital interactions are taken into account. This result, from the point of view of the phenomenology of the Mott insulators in spin-orbit coupled $f$-electron systems, suggests that topological phases should be sought near the metal/insulator transition in   systems where the ligand band is derived from, e.g.,  $d$ orbitals, for which correlation effects could be non-negligible. 

\begin{figure}  
    \centering
    \includegraphics[width=0.45\textwidth]{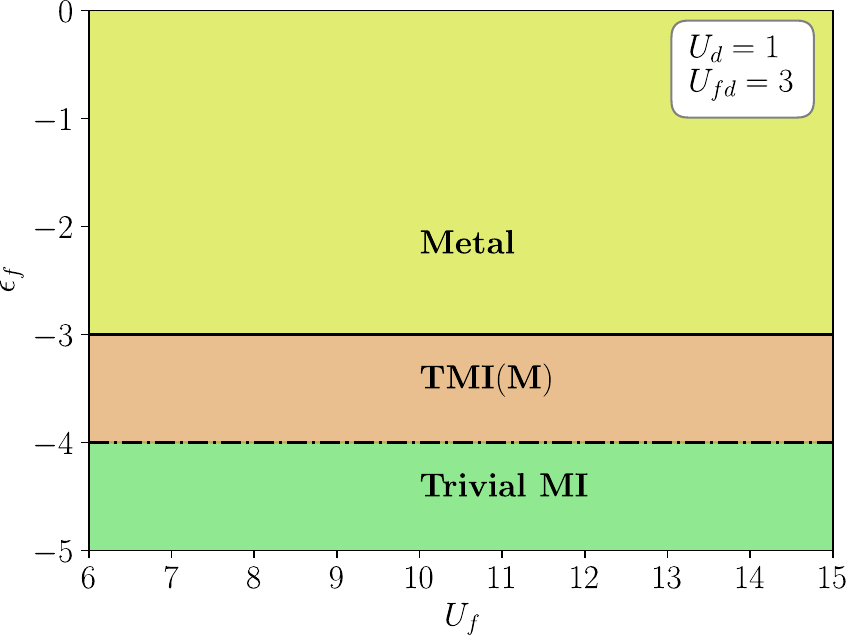}
    \caption{Phase diagram on the $\epsilon_f$-$U_f$ plane for the system with filling 1. Dashed-dot and solid lines mark the metal/insulator and topological transition, respectively. Non-zero $U_d$ and $U_{fd}$  stabilize   the topological Mott insulator phase in a region near the Mott transition. Here $U_d=1$, $U_{fd}=3$, and $V=0.5$.}
    \label{fig3}
\end{figure}

Moreover, we have found that topological phase transitions are not associated with a spectral gap closing in accord with the phenomenon called {\it first-order topological phase transition}. This result is analogous to a recent finding \cite{Wysokinski2023} of a similar behavior at the phase transition between the quantum anomalous Hall insulator and the trivial insulator. Here, we have shown that the effect also takes place at transitions between two different topologically non-trivial phases   suggesting the universality of the absence of the spectral gap closing in systems dominated by Mott physics. We show that instead of spectral gap closing the topological transitions are signaled by a kink in the spectral function appearing  at one of the time reversal invariant momenta. Combining this result with the findings of Ref.~[\onlinecite{Wysokinski2023}] we suggest that kinks or in general singularities in the spectral functions at time reversal invariant momenta are universal markers of topological phase transitions in systems dominated by Mott physics.

 \section*{Acknowledgements}
The work is supported by the Foundation for Polish Science through the International Research Agendas program co-financed by the European Union within
the Smart Growth Operational Programme. 
KB also acknowledges   the support of the  Deutsche Forschungsgemeinschaft under the {\it Transregional Collaborative Research Center} TRR360. W.B. acknowledges support by Narodowe Centrum Nauki (NCN, National Science Centre, Poland) Project
No. 2019/34/E/ST3/00404. 

\appendix
\section {Phase diagram for filling 1}

All general results and conclusions for the system with filling 1 are the same as  for the case with filling 3. These include an absence of the spectral gap closing and associated appearance of kinks in the spectral functions at the topological transitions and the presence of topological phase on the verge of the metal/Mott insulator transition only in the presence of non-zero $U_d$ and $U_{fd}$.  For the sake of completeness in Fig.~\ref{fig3} we present a phase diagram for $U_d=1$ and $U_{fd}=3$ (parameters are the same as   in Fig.~\ref{fig0}).

\end{document}